\documentclass[runningheads]{llncs}

\usepackage[T1]{fontenc}
\usepackage{graphicx}
\usepackage{amsmath}
\usepackage{amssymb}
\usepackage{amsfonts}
\usepackage{physics}
\usepackage{amssymb}
\usepackage{verbatim}

\begin{document}
\title{Evolutionary reduction of the laser noise impact on quantum gates}
\date{April 2023}



\author{
    Tam'si Ley\inst{1, 3} \orcidID{0009-0001-5881-5689} \and
    Anna Ouskova Leonteva\inst{1} \and
    Johannes Schachenmayer\inst{2} \and
    Pierre Collet \inst{1}
}

\authorrunning{T. Ley et al.}

\institute{
    ICUBE, Université de Strasbourg \and
    ISIS, Université de Strasbourg \and
    QuantFi, Paris
}

\maketitle              
\noindent
\makebox[\linewidth]{
April 2023}

\begin{abstract}


As the size of quantum hardware progressively increases, the conjectured computational advantages of quantum technologies tend to be threatened by noise, which randomly corrupts the design of quantum logical gates. Several methods already exist to reduce the impacts of noise on that matter. However, a reliable and user-friendly one to reduce the noise impact has not been presented yet. Addressing this issue, this paper proposes a relevant method based on evolutionary optimisation and modulation of the gate design. This method consists of two parts : a model of quantum gate design with time-dependent noise terms, parameterised by a vector of laser phases, and an evolutionary optimisation platform aimed at satisfying a trade-off between the gate fidelity and a pulse duration-related metric of the time consuming simulation model. This feature is the main novelty of this work. Another advantage is the ability to treat any noise spectrum, regardless of its characteristics (e.g., variance, frequency range, etc). A thorough validation of the method is presented, which is based on empirical averaging of random gate trajectories. It is shown that evolutionary based method is successfully applied for noise mitigation. It is expected that the proposed method will help designing more and more noise-resisting quantum gates. 


\end{abstract}

\section{Introduction}
    Quantum computing uses the laws of quantum physics to perform logical operations inaccessible to classical means: e.g., superposition and entanglement. In theory, an ideal quantum computer would solve some mathematical problems with exponentially fewer computing resources than any classical counterpart \cite{boson}. It applies to practical fields such as quantum chemistry \cite{chem}, materials science \cite{mater}, and cryptography \cite{num}.

However, quantum systems are highly prone to a noise, coming from interactions with the environment, or from imperfections in the control apparatus (e.g., laser, electronic devices). Due to analog nature of those quantum systems, the noise critically affects the performance of quantum computers. Moreover, the noise also dramatically reduces the quantum computational power, and would prevent quantum computers from reaching any advantage on practical problems with regard to the classical counterparts. This can be seen by the following result: assuming a constant error rate and without any effective correction strategy, a large variety of random quantum circuits (i.e programs) are efficiently simulated by classical computers \cite{sim1}, \cite{sim2}, \cite{sim3}, which would be far more difficult without noise. The ability to overcome noise is the central question raised by quantum computing skeptics \cite{skep}. Indeed, quantum circuits (programs) solving relevant problems, require millions of elementary  quantum gates (logical operators). Thus, even with small noise rates, they need efficient strategies to suppress errors. The latter exist in theory, but they require average gate infidelity much lower than what current noise levels impose. Reducing the impact of noise on the gate design is a highly non-trivial challenge, for which several solutions have been proposed: e.g., \cite{ff1}, \cite{ff2}, \cite{ff3}. Generally, these methods model the gate infidelity (inaccuracy) with equations that require long simulations, and do not cope well with a high standard deviation or high frequency fluctuations in the noise.
\newline \hphantom{M}
Addressing these issues, in this work, we focus on adapting a flat laser pulse suffering amplitude and dephasing noises, in order to implement a Rydberg blockade two-qubit gate, which is specific to the cold-atoms hardware. The theoretical information about that hardware is largely presented in the literature: e.g., \cite{Stras}. In order to reduce the noise and simultaneously, to reduce the time spent in the Rydberg state, we formulate a quantum gate design as a continuous non-constrained bi-objective problem based on a time-expensive computer simulation.
Taking into account the number of objectives ($2$) and the large number of decision variables ($51$), we set 100 candidate solutions in the population, where a computation time of around 1 minutes per each solution. In order to reduce the computational complexity, we use the CPU-parallel celebrated multi-objective algorithm Non-dominated Sorting Genetic Algorithm III (NSGA-III) provided by the EASEA ({\em EAsy Specification of Evolutionary Algorithms}) platform described in \cite{easea}. \newline \hphantom{M}
To our best knowledge, this work is the first attempt to present a quantum gate design task as a continuous large-scale multi-objective problem, solved by parallel optimisation algorithm. Our method is not limited by any characteristic of the noise, and allows us to select a pulse sequence (laser phases) and duration to implement the target gate. Moreover, EASEA provides different single and multi-objective optimisation algorithms, which can be easily applied for the problem of laser noise reducing in the further researches in order to investigate an impact of different optimisation techniques on the performance in terms of the accuracy and speed up.
\newline \hphantom{M}
This paper is organized as follows. Section 2 briefly outlines the related works. Section 3 describes the problem definition. Section 4 presents the proposed method based on the evolutionary optimisation. The experiments and results are shown in Section 5. Section 6 concludes this article.
\section{Related works} \vspace{-1.5 pt}
    \hphantom{M} In this section, we briefly present the related works addressing the problem of laser noise impact reduction for quantum gate design presented in the literature. They can be broadly divided in four categories.

The first one is called noise mitigation, and consists of repeating some gates throughout the circuit, in order to reduce the influence of noise on the computation result \cite{nm1}. This technique generally assumes that the gate noise is uncorrelated in time. However, correlated noise has been shown to bypass the current noise mitigation strategies, as experimentally proved in \cite{nm2}. Thus, non-trivial correlations in time must be accounted for, when analysing noise impact on quantum computation. 


The second category is robust optimal control, as was outlined in \cite{roc}. It converts noise impact minimisation, into an optimal control problem in an extended space with appropriate boundary conditions, and derives semi-analytical solutions to that problem. However, it only treats time-independent noise in a single direction (e.g amplitude or dephasing). Adapting that technique to time-dependent noise would increase the problem dimension exponentially in the number of harmonics, when considering all the cross products between operators. A slight variation of this approach is "inverse geometric optimisation", as introduced in \cite{igo}. The idea is to convert noise impact reduction into a Euler-Lagrange problem, leading to a second order non-linear differential equation. This approach allows to directly track the trajectory of the quantum state in a semi-analytical formulation. However, this method cannot be currently be adapted to either dephasing and/or time-dependent noise, due do its mathematical formulation \cite{igo}.

The third category is the filter function formalism, as outlined in \cite{ff1} \cite{ff2} \cite{ff3}. The idea is to formulate the average gate infidelity as a function of both the laser controls and the noise power spectral density. Then, it is optimised with differentiable methods such as gradient descent. However, that method fails when the noise intensity is of order $10\%$ of the laser pulse intensity, such that the implicit first-order truncation does not ensure a low enough infidelity.

The fourth category consists of estimating the noise impact with Monte Carlo summation of random gate trajectories; and to optimise the laser pulse with a separate algorithm. In \cite{fourier}, the laser control is decomposed in Fourier series, and the Fourier coefficients are optimized through the Nelder-Mead algorithm. In their work, dephasing noise is considered relative, like amplitude noise. This stands in contradiction to the convention in the literature. Our present work follows a similar path, but characterises the laser control with flat pulses, whose phases are optimised by a multi-objective algorithm.
\section{Problem definition}
\subsection{Context}
    Optically-trapped cold atoms constitute a promising hardware architecture for long term quantum computing due to long qubit life time, maturity of laser technology and flexibility of two-qubit interactions \cite{Stras}. The idea of this technology is to encode bits as two specific orbitals, called "0" and "1", populated by the valence electron of some chosen special atomic species \cite{Stras}. 
    \begin{figure}[h]
    \centering
    \includegraphics[height=2.8cm]{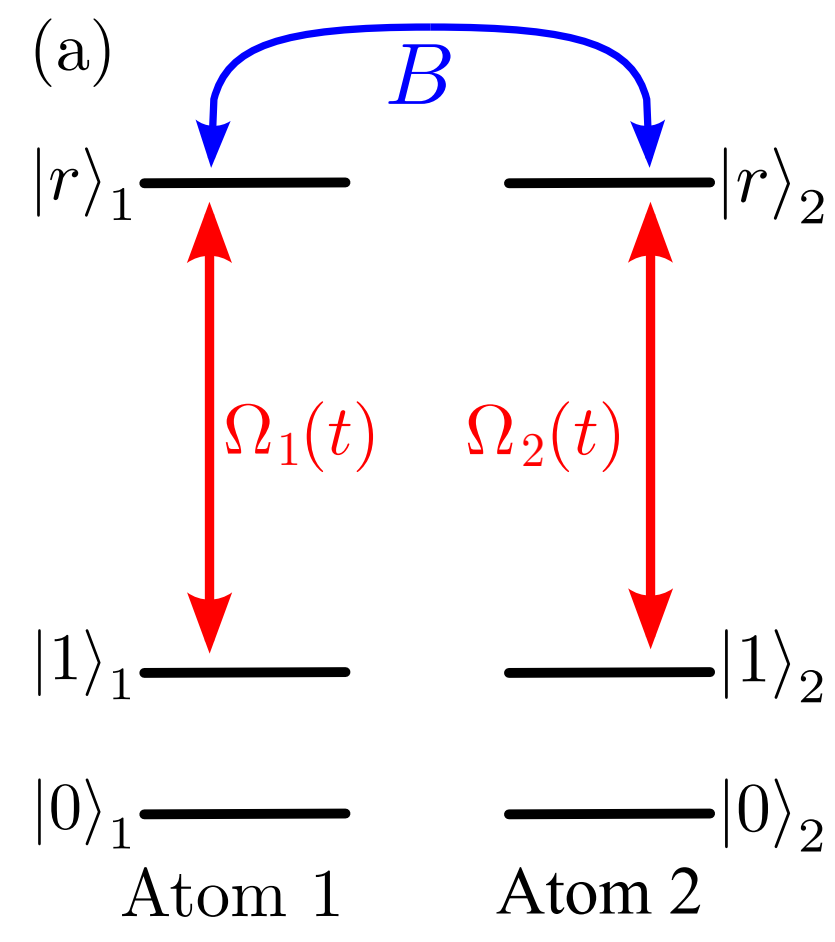}
    \caption{Atomic transitions and Rydberg interaction. Illustration adapted from \cite{sjan}.}
    \label{figAtom}
    \end{figure} \vspace{-2em}



  
    Upon interaction with a nearby electromagnetic field, electrons switch between orbitals. This is usually done with laser tuned in resonance with the atomic transition frequency between the said orbitals. In the literature, a $\pi$-pulse refers to a complete transfer between two orbitals, with a multiplication by the imaginary unit of the relevant orbital coordinate in the quantum state vector.
    Rydberg states refer to orbitals far away from the nucleus, exhibiting high electric dipole moments. When the resulting electric potential is much larger than the power of the electromagnetic field, the Rydberg blockade prevents two electrons to be simultaneously in a Rydberg state. This happens when atoms are closed to each other. The protocol outlined in \cite{block} allows to design a universal two-qubit gate, using only the Rydberg blockade and four $\pi$-pulses between the "1" and a Rydberg orbitals. Thus, as we focus on laser noise, we only have to optimise the $\pi$-pulse. Figure~\ref{figAtom} shows a simplified view of the atomic transition described above, where $r$ is a Rydberg state, $B$ is the Rydberg blockade strength and $\Omega_j$ is a function characterising the laser pulse. \vspace{-1em}

\subsection{Mathematical model}

    Due to the mathematical description of quantum mechanics, the electron's quantum state is represented as a complex unit vector, whose dimension is the number of considered orbitals. In our case, there are 2 : the $1$ and $r$ orbitals, as $0$ is excluded from the transitions. Since the evolution of the vector is linear and norm-preserving, it is represented as a $2 \times 2$ complex unitary matrix $U(t)$, with $U(0)$ being the identity matrix. The interaction between the atom and an electromagnetic field, inducing the orbital transition, is mathematically represented as the following differential equation:
    \begin{equation} \label{eq:iter}
        \dot{U}(t) = \imath\hat{H}(t)U(t)
    \end{equation}
    where $\hat{H}(t)$ is the hamiltonian matrix, which corresponds to the energy configuration of the system.
    The transition between two atomic orbitals is characterised by the complex Rabi frequency $\Omega$, as shown in the equation below:
    \begin{equation} \label{eq:trans}
        \hat{H}(t) = 
        \begin{pmatrix}
            \varepsilon_d(t) & (1+\varepsilon_a(t)) \Omega^*(t) \\
            (1+\varepsilon_a(t)) \Omega(t) & -\varepsilon_d(t)
        \end{pmatrix}
    \end{equation}
    where $\varepsilon_d$ is the dephasing noise, $\varepsilon_a$ is the amplitude noise. In the absence of noise, the pulse will be the fastest, if the Rabi frequency has a constant modulus, as explained in \cite{sjan}. Since we want the gates to be fastest as possible, in order to avoid other types of noises, we accordingly parameterise the Rabi frequency :
    \begin{equation}
        \Omega = 2\pi f_{\max} e^{\imath \varphi}
    \end{equation}
    where $f_{\max}$ is the maximal frequency of the laser pulses and $\varphi$ is the phases of the laser pulses. In the noiseless case, the phase is constant $\varphi = 0$. Here, we set it as a piece-wise constant function, with a fixed number of slices ($N$): $\varphi(t) = \varphi_i$ for $t \in [\frac{iT}{N}, \frac{(i+1)T}{N}]$. The solution of the differential Equation~\ref{eq:iter} is a random matrix written as $U(\varphi, t)$.
    For the sake of clarity, the list of all the mathematical symbols used in this work is summarised in Table~\ref{tab:symbol}. 
    \begin{table}[h]
        \centering
        \begin{tabular}{|c|c|}
            \hline Symbol & Significance (unit)\\
            \hline $\ket{0}$ & Ground orbital (--) \\
            \hline $\ket{1}$ & Excited orbital (--) \\
            \hline $\ket{r}$ & Rydberg orbital (--) \\
            \hline $\imath$ & Complex imaginary unit (--) \\
            \hline $t$ & Running time (1/$f_{\max}$) \\
            \hline $T$ & Pulse duration (1/$f_{\max}$) \\
            \hline $U$ & Pulse matrix (--) \\
            \hline $\dot{U}$ & Derivative of pulse matrix (--)\\
            \hline $\Omega$ & Rabi frequency (rad $f_{\max}$) \\
            \hline $\Omega^*$ & Complex conjugate thereof (rad $f_{\max}$)\\
            \hline $\hat{H}$ & Hamiltonian matrix (rad $f_{\max}$) \\
            \hline $U_0$ & Target pulse matrix (--) \\
            \hline $f_{\max}$ & Maximal Rabi frequency (1)\\
            \hline $\nu_0$ & Maximal noise frequency ($f_{\max}$) \\
            \hline $\boldsymbol{\varphi}$ & Vector of laser phases (rad)\\
            \hline $N$ & Number of time slices/laser phases \\
            \hline $F$ & Pulse infidelity (-)\\
            \hline $G$ & Time in Rydberg state (1/$f_{\max}$)\\
            \hline $\varepsilon_a$ & Amplitude noise (rad $f_{\max}$)\\
            \hline $\varepsilon_d$ & Detuning noise (rad $f_{\max}$)\\
            \hline
        \end{tabular}
        \caption{List of symbols used in the problem definition}
      \label{tab:symbol}
    \end{table}
    \vspace{-3em}
    This model can be transformed to an optimisation problem. From an optimization point of view, the real-world quantum gate design problem, presented above, is a typical expensive continuous bi-objective optimization problem without constraints. A bi-objective problem aims at optimizing two values of interest (objectives), which are antagonist. The answer is a set of trade-offs solutions (a vector with the input parameters) between the defined objectives. In section 3.3, we present its inputs and objectives. \vspace{-1em}


\subsection{Values of interest and input parameters}
    
    In this study, we consider 2 values of interests (objective functions) and 51 input parameters. \vspace{-1em}
    
    \subsubsection{Input parameters:}
    
        The parameters and their ranges of variations are summed up in Table~\ref{tab:input}, 
        \begin{table}
        \centering
        \caption{List of all inputs}\label{tab:input}
        \begin{tabular}{|l|l|l|l|}
        \hline
        Input &  Min  & Max & Unit\\ 
        \hline
        Phases: $\boldsymbol{\varphi}[0-50]$ &  0 & 2$\pi$ & rad \\
        Time: $T$ & 1 & 5 & $1/f_{\max}$\\
        \hline
        \end{tabular}
        \end{table}
\vspace{-2em}        
        where the following notation is used: \vspace{-1em}
        \begin{itemize}
            \item $\boldsymbol{\varphi}$ is a vector of $N$ elements, which refers to the values of the laser phase in the orbital transitions. In the current model $N = 50$ as a reasonable number required for both processes: simulation and optimisation.
            \item $T$ is the duration of the model simulation, which has an impact on the time spent by the electron in the Rydberg state ($G$); 
            \item $f_{\max}$ is the maximal frequency the laser emits to. In our problem definition, we normalise all units with $f_{\max} = 1$, which means that the model can be applied to any value of that frequency. In recent experiments, typically $f_{\max} = 1$ GHz.
        \end{itemize}
    \vspace{-1em}    
    \subsubsection{Objective functions}
    
    We seek to make $U(\boldsymbol{\varphi}, T)$ (see Equation 1) as close as possible to $U_0 := \begin{pmatrix} 0 & \imath \\ \imath & 0 \end{pmatrix}$, because this matrix represents the $\pi$-pulse that we look for, since we base our work on the protocol of \cite{block}, which features only $\pi$-pulses. This motivates us to define the gate infidelity as follows :
    \begin{equation}
        F(\boldsymbol{\varphi}, T) := \mathbb{E}(1 - \left|\text{Tr}(U(\boldsymbol{\varphi}, T)U_0^\dagger)\right|^2/4)
    \end{equation}
    A important source of errors in quantum computation, is spontaneous photon emission. In this random process, super excited states like the Rydberg ones, tend to spontaneously revert to a lower-energy state, releasing a photon. This cannot be prevented, so the only way to minimise the risk of occurrence is to minimise the time spent by the electron in the Rydberg orbital. Assuming the electron starts in the $1$ orbital, this quantity is given by :
    
     \begin{equation}
        G(\boldsymbol{\varphi}, T) := \mathbb{E}\left(\int_0^T | \langle r | U(\boldsymbol{\varphi}, t) |1\rangle |^2 dt\right)    
   \end{equation}
    
    
    Let's briefly show that these two quality criteria are antagonistic. Simulating flat pulses leads to the appearance of noise terms of the form $\varepsilon(\frac{jT}{N})$ in the computations.
    Since the number $N$ of slices is fixed, these terms tend to vanish as $T$ grows to infinity. Thus, the impact of noise can be mitigated by extending the pulse duration, and choosing a trivial $\boldsymbol{\varphi}$. In other words, increasing $T$ tends to reduce $F$, as well as simplifying the choice of $\boldsymbol{\varphi}$, but it would potentially imply a larger value of $G$ and thus a higher risk of Rydberg state decay. On the other hand, $0 \leq G \leq T$ (because the integrand is lower than $1$), so an optimal $G$ would rather be found by decreasing $T$, but in turn this complexifies the search for optimal phases. Consequently, to find the trade-off between these objectives is required. \vspace{-1em}






    

\section{Proposed Method}

The goal of this study is to reduce the gate infidelity (i.e the impact of laser noise) by finding the optimal laser phases $\boldsymbol{\varphi}$, without increasing the time spent in the Rydberg state, in order to minimise the risk of Rydberg state decay. Consequently, it will improve the global gate quality. To achieve this goal, we propose a CPU-parallel multi-objective optimisation-based method, since the quantum gate design can be formulated as a continuous bi-objective computationally expensive problem (as was shown above). Indeed, a multi-objective optimization process allows us to find: (i) all possible non dominated solutions -- combinations of $\boldsymbol{\varphi}$ and $T$, which is called the Pareto set; (ii) its image in the objective space, which presents the Pareto front. Thus, we consider to define a set of optimal parameters of the model (i.e., the phases vector of the laser pulses and the duration of the model simulation), by minimizing the objective functions (i.e., overall infidelity and time spent in the Rydberg state), calculated by simulation.
\begin{figure}[h]
    \centering
    \includegraphics[height=6cm]{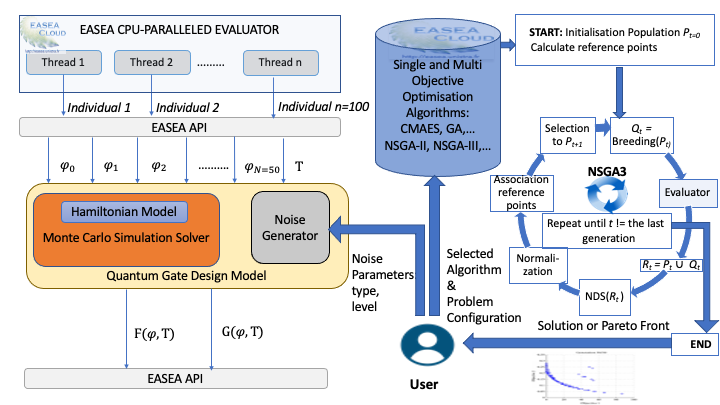}
    \caption{Simplified scheme of the proposed method.}
    \label{figMethod}
\end{figure}
The proposed method consists of two following parts:
\begin{itemize}
\item A Quantum gate design presented in Section 3.2, which consists of simulating the laser pulse hamiltonian, on noisy trajectories; and computing the objective functions with Monte Carlo average. For the sake of completeness, the noise generation details are provided in online supplementary materials: \url{https://git.unistra.fr/tsley/thesis}.
\item The EASEA platform, an open-source compiler, which supports single and multi-objective optimization thanks to its own base of the templates of different state-of-the-art algorithms \cite{easea}. The main advantage of EASEA is that it allows scientists to apply evolutionary algorithms to solve their real-world problems through a user-friendly interface with CPU and/or GPU parallelisation. More precisely, this compiler automatically couples one of the algorithm template files and a problem to be optimised, and provides the C++ based code of the selected algorithm with the integrated problem in it and the compiling file. Consequently, it suits perfectly as a software support for our study. Notably, the quantum gate design model can be modified without changing other parties of the code, due to the very flexible structure of EASEA.
\end{itemize}

By this concept, we had to select an optimisation algorithm template (\textit{*.tpl} file), define the problem in terms of the pseudo-code (\textit{*.ez} file) and define the noise parameters, such as the spectrum (white, pink, etc.) and its level w.r.t. maximal intensity of the laser pulse. The simplified scheme of the considered method is schematically depicted in Figure~\ref{figMethod}. It details the programming relation between: (i) the EASEA CPU-parallel evaluator with our quantum gate design model presented on the left and (ii) the selected multi-objective algorithm:  Non-dominated Sorting Genetic Algorithm (NSGA-III), shown on the right in Figure~\ref{figMethod}. NSGA-III originally presented in \cite{nsga3}, is based on the non-dominated sorting selection mechanism and uses a predefined set of reference points to ensure diversity in the solutions \cite{nsga3}. NSGA-III was selected in this work between other templates, because of the following reasons: (i) it is efficient on several multi- and many-objective problems \cite{unsga3} (in future, the current problem can be extended to a many-objective version); (ii) it does not require any additional hyper-parameters;  (iii) it introduces a computationally fast approach, by which the reference point set is adaptively updated on the fly based on the association status of each reference point over a number of generations \cite{nsga3}, \cite{unsga3}; (iv) it can handle constraints without introducing any new parameter, which can be useful in our next research, where some constraints will be taken into account; (v) it can easily be scaled for solving single-objective problems \cite{unsga3}. The following steps explain how this method performs:
\begin{enumerate}
    \item The selected optimization algorithm, e.g., NSGA-III, starts by randomly generating the initial set of solutions, where each solution consists of the parameters presented in Table~\ref{tab:input};
    \item Its main cycle 
    is repeatedly executed until the last generation (see the right part of Figure~\ref{figMethod});
    \item NSGA-III creates the new child population $Q_t$ from the parent population $P_t$ by the classical breeding operators used in NSGA-III \cite{nsga3}: Simulated Binary crossover (SBX) crossover and Polynomial mutation operator. 
    \item The \textit{EASEA Evaluator} executes 100 threads to compute $F$ and $G$ in parallel (see the left part of Figure~\ref{figMethod}). Each executed thread works out the model code with its own combination of input parameters and generated noise.
    \item Each thread returns the output values (the overall infidelity ($F$) and the time spent in the Rydberg state ($G$)) back to the NSGA-III.
    \item NSGA-III performs according to its description presented in \cite{nsga3} and goes to the next generation.
\end{enumerate}

Also, important to emphasise that  NSGA-III of EASEA is designed for CPU parallel use (several hundred  computing threads), which allows us to reduce the runtime. \vspace{-1em}

\section{Experimental analysis}
    \subsection{Experimental design}

    All values of the setting parameters (both for the simulation model and for the optimisation algorithm) presented below, are the same during all provided experiments.  For all experiments we use the problem definition presented in Section 3.3. The decision variables are defined as the presented input parameters. For single-objective study, we use only $F$ as a single optimisation function and applied the well-known Covariance Matrix Adaptation Evolution Strategy  (CMA-ES) algorithm \cite{cmaes} from the EASEA plateform. For multi-objective study, as it was presented above, the NSGA-III algorithm is applied. \vspace{-1em}

    \subsubsection{Model Settings}
        
        In this paper, we chose to work with $1/f$ noise, also known as pink or flicker noise, since it is widely present in electronic devices, and thus in the quantum hardware that they control.
        So, we generate $10^6$ samples thereof, each time as a sum of sines and cosines with $25$ uniformly log-spaced harmonics, in the interval $[f_{\max}, 100f_{\max}]$, thereby approximating a high-frequency noise. Remind that $f_{\max}$ is the characteristic unit of the problem, and is set to $1$. 
        Remind that the optimisation target is a $\pi$-pulse to implement a Rydberg blockade two-qubit CZ gate. We effectuate three tests with respective noise levels of 10\%, 20\% and 30\% of the maximal laser pulse intensity.
    \vspace{-1em}
    \subsubsection{Optimisation Settings}
    
        For all experiments, the number of decision variables equals 51, according to the problem definition. The parameter settings of CMA-ES is limited by the number of generations = 300, the population size = 100 and the number of threads = 100. 
           \begin{table}[h]
        \centering
        \caption{ Parameter settings of NSGA-III}\label{tab:nsga3-set}
        \begin{tabular}{|l|l|l|l|l|l|l|l|l|l|l|}
        \hline
        Parameter  & $p_c$ & $\eta_c$ & $p_m$ &  $\eta_m$&$M$& $P$ & $H$  & $N$ & $G$ & $Th$ \\ 
        \hline
        Value & 1.0 & 30& 1/51 & 20& 2& 99& 100 & 100 & 200& 100\\
        \hline
        \end{tabular}
        \end{table} 
    The parameter settings of NSGA-III are presented in Table~\ref{tab:nsga3-set}, where $p_c$ is Simulated Binary (SBX) crossover probability, $\eta_c$ is crossover distribution index,  $p_m$ is polynomial mutation probability, $\eta_m$ is polynomial distribution index, $M$ is the number of objectives, $P$ is a given integer value, which refers to the number of divisions considered along of each objective axis, $H$ is the number of reference points, $N$ is the population size, $G$ is the number of generation and $Th$ is the number of threads. The selection of $P$, $H$ and $N$ was made according to the rules defined in \cite{nsga3}: (i) if a problem has less number of objectives, a larger number of $P$ is required (100 is recommended value in bi-objective case); (ii) $H$ is calculated by the approach of Das and Dennis \cite{ref-points}; (iii) the population size is set as the smallest multiple of four higher than the number of reference points (in our case $N$=$H$).     
   
\subsection{Result analysis}
\begin{figure}[h]
        \centering
        \begin{tabular}{cc}
        \includegraphics[height=2.6cm]{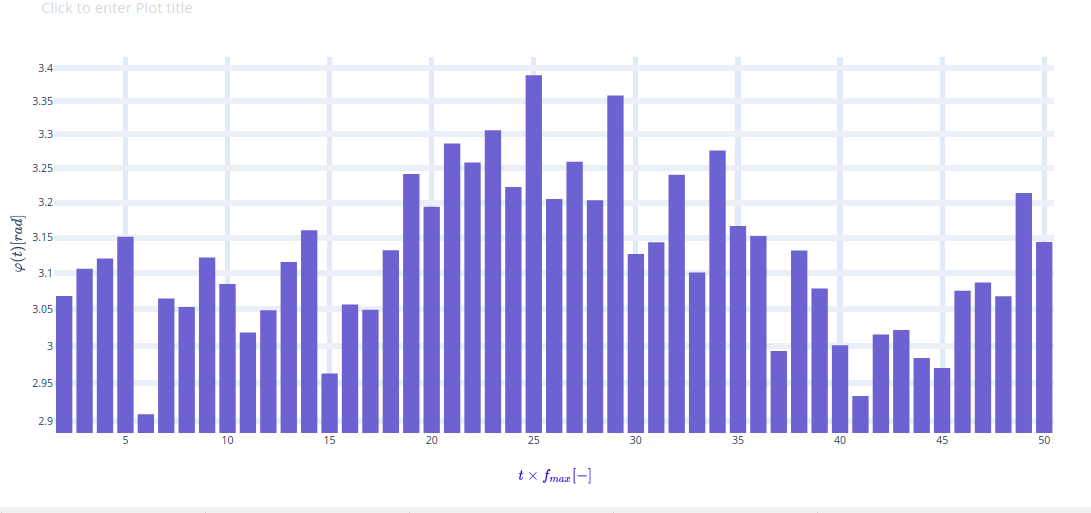}&
        \includegraphics[height=3cm]{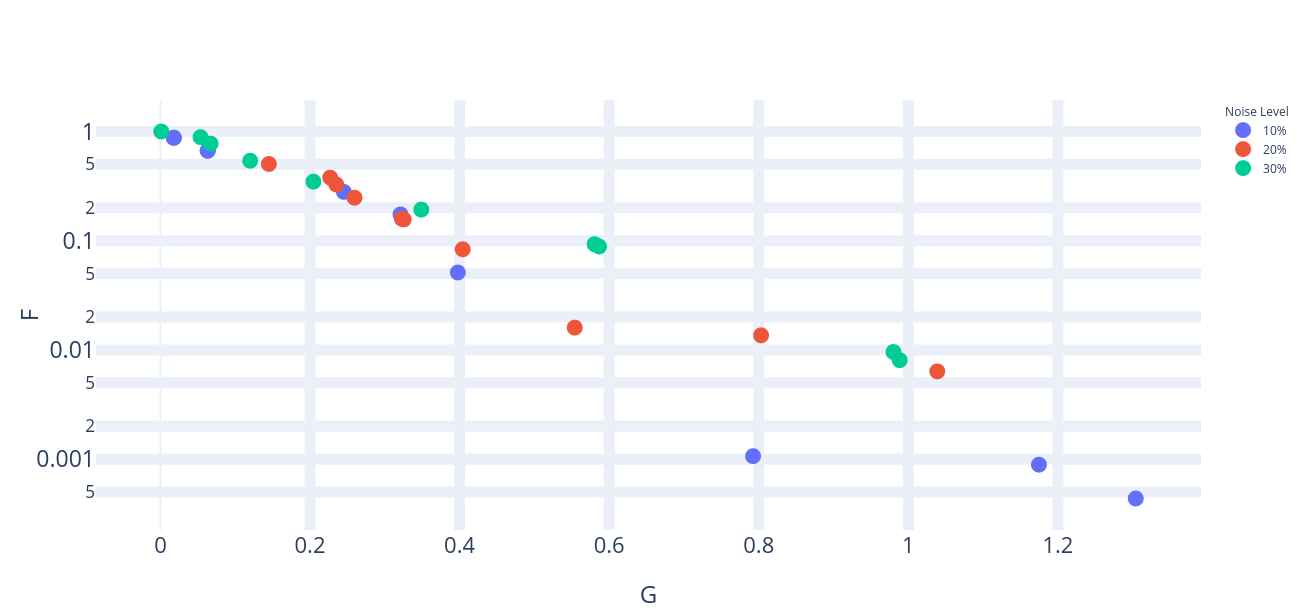} \\
        a) Best solution& b) Pareto Front
        \end{tabular}
        \caption{Obtained results.}
        \label{fig:pareto}
    \end{figure}

    The obtained results are shown in Figure~\ref{fig:pareto} and described below.
    {}
    \vspace{-3em}
    \subsubsection{Single-objective}
        This study started with a single-objective optimisation of $F$, with a fixed $T=1$ and a noise intensity of 10\%. After 136 generations of 100 candidates, the algorithm CMA-ES found an optimal solution $F = 0.005052$. This solution in terms of the coordinates of $\boldsymbol{\varphi}$ is presented in log scale in Figure~\ref{fig:pareto} (a). This result is better than the gate infidelity $F = 0.0055$ obtained with the naive choice of setting all the 50 phase to be zero: $\boldsymbol{\varphi} = \bold{0}$. This result confirmed us that artificial evolution can bring some improvements in robust gate designs, and motivated us to explore multi-criteria optimisation for simultaneous antagonistic problems. 
    \vspace{-1em}
    \subsubsection{Multi-objective}
    

        The Pareto fronts obtained for three selected noise levels (\%10, \%20 and \%30 of $2\pi f_{\max}$) are presented in Figure~\ref{fig:pareto} (b), where the axe $F$ is presented in log scale. As it seen from Figure~\ref{fig:pareto} (b), the lowest minimum of $F$ was obtained in the case of 10\% noise level (presented in blue).   
 
        For the rightmost element of the Pareto front, the pulses we find result in a compromise between an infidelity slightly higher than the naive case, and a shorter time spent in the Rydberg state, signalling a preference in reducing Rydberg decay risk. For elements of the leftmost Pareto front, the situation is reversed. This shows that NSGA-III drives the solutions to satisfactory compromises, which should be investigated w.r.t. possible values of the Rydberg decay rate. For all cases, the shape of the Pareto Front is slightly discontinuous, however the solutions are not irregularly distributed in the objective space. It shows that for all selected noise levels, NSGA-III is able to find solutions in the different regions of the objective space, providing the possible variations of trade-off between the overall infidelity and the time spent in the Rydberg state. We assume that these results can be improved by increasing the population size and the number of generations. \vspace{-1em}
\section{Conclusion}
    The multi-objective based method we propose aims at introducing a reliable gate design, which allows us to reduce both the laser noise impact and the risk of Rydberg state decay. As a result, this method provides a variety of trade-offs between these two objectives, within a reasonable run-time, thanks to CPU-parallelization. It can be easily applied to different noise and laser pulses models. The problem studied here focuses on decomposing a flat pulse affected with pink noise on both amplitude and phase with different noise intensity levels. With the trust of the multi-objective optimization method, thanks to EASEA platform, it is now possible to look for interesting gate design features for a wide variety of noise with a reduced run-time. Moreover, as the EASEA platform provides different single and multi-objective optimisation algorithms, which can be easily coupled with the problem of laser noise reduction, we prospect to investigate the impact of different optimisation techniques on the performance in terms of the accuracy and speed up in further works. Also, further research with different noise types would be considered.
However, our work inherited the drawbacks of the protocol outlined in \cite{block}, in the sense that its setup would require an individual laser for each atom, which is difficult to attain. Therefore, beyond simply noise-corrected $\pi$-pulses, future works will involve a global pulse model with higher degrees of freedom,
similarly to \cite{sjan}. This would correspond to a single laser for all the atoms, able to design multi-qubit Rydberg blockade gates. It is expected that this method can help to make a qualitative step toward developing robust quantum gates. \vspace{-1em}


%
%
%
%

\newpage
\section{Supplementary material}
    \subsection{Symbols table}
    \begin{figure}[h]
        \centering
        \begin{tabular}{cc}
        \end{tabular}
    \end{figure}
    \begin{table}[]
        \centering
        \begin{tabular}{|c|c|c|}
            \hline Symbol & Significance & Dimension (Unit) \\
            \hline $\ket{0}$ & Ground orbital & Dimensionless \\
            \hline $\ket{1}$ & Excited orbital & Dimensionless \\
            \hline $\ket{r}$ & Rydberg orbital & Dimensionless \\
            \hline $\imath$ & Complex imaginary unit & Dimensionless \\
            \hline $t$ & Running time & Time (1/$f_{\max}$) \\
            \hline $T$ & Pulse duration & Time (1/$f_{\max}$) \\
            \hline $U$ & Pulse matrix & Dimensionless \\
            \hline $\dot{U}$ & Derivative of pulse matrix & Dimensionless \\
            \hline $\Omega$ & Rabi frequency & Angle frequency (rad $\cdot$ Hz) \\
            \hline $\Omega^*$ & Complex conjugate of the latter & Angle frequency (rad $\cdot$ Hz) \\
            \hline $\hat{H}$ & Hamiltonian matrix & Dimensionless \\
            \hline $U_0$ & Target pulse matrix & Dimensionless \\
            \hline $\nu_0$ & Maximal value of the Rabi frequency & Inverse time (Hz) \\
            \hline $f_{\max}$ & Maximal frequency reached by the noise & Inverse time (Hz) \\
            \hline $\varphi$ & Laser phase & Angle (rad) \\
            \hline $F$ & Pulse infidelity & Dimensionless \\
            \hline $G$ & Time in Rydberg state & Time (1/$f_{\max}$) \\
            \hline $\varepsilon_a$ & Amplitude noise & Angle frequency (rad Hz) \\
            \hline $\varepsilon_d$ & Detuning noise & Angle frequency (rad Hz) \\
            \hline
        \end{tabular}
        \caption{List of symbols}
        \label{tab:my_label}
    \end{table}

\subsection{Noise generation}

    Pink noise is also referred to as "$1/f$ noise", since its power spectral distribution is roughly the inverse function, though with a minimal and maximal frequency cutoffs, since the inverse function is not integrable.
    Being also named "flicker" noise, it is widely present in electronics, including the ones that control quantum devices. Existing methods to compute pink noise with trigonometric series present a lot of limitations : for example they require a large number of harmonics to ensure convergence to the desired spectrum, and thus are not presented explicitly in the literature. The following method accurately computes pink Noise with a limited number of terms. At the best of our knowledge, it has not been presented elsewhere.
    
    $$
        C(h)
        = \int_{0}^{\log(\nu_{\max})} \cos(2\pi e^x h) dx
        = \log(\nu_0) \int_{0}^{1} \cos(2\pi \nu_0^s h) ds
    $$
    Indeed, this integral can be approximated up to first order as :
    $$
        \frac{C(h)}{C(0)} = C_M(h) + O\left(\frac{1}{M}\right)    
    $$
    Where
    $$
         \quad C_M(h) := \frac{1}{M} \sum_{n=0}^{M-1} \cos(2\pi \nu_0^{\frac{n}{M}} h)
    $$
    Thus, if $(a_n, b_n)$ are two i.i.d sequences of standard normal random variables, then the expression
    
    $$
        X(t) := \frac{1}{M} \sum_{n=0}^{M-1} \cos(2\pi \nu_{\max}^{n/M} t) a_n + \sin(2\pi \nu_{\max}^{n/M} t) b_n
    $$
    defines a Gaussian process such that $\mathbb{E}(X(t+h)X(t)) = C_M(h)$, thereby providing a good approximation to pink noise.

\subsection{Optimal coordinates of $\varphi$}

    The phase vector found by the single-objective optimisation mentioned in subsection 5.2 is :
    \begin{align*}
    \varphi = (\\
        &3.1117210566,  3.0681531508,  3.1062684923,  3.1204162648,  3.1511899718,\\ &2.9093600733,  3.0648268235,  3.0531992808, 3.1219826294,  3.0848634591,\\ &3.0181634846,  3.0485478461,  3.1157769830,  3.1606018942,  2.9632053356,\\ &3.0567039551, 3.0493017295,  3.1321501012,  3.2415897692,  3.1940974328,\\ &3.2861696576,  3.2583179501,  3.3060153192,  3.2227544442,  3.3887061970,\\ &3.2053472380,  3.2594614180,  3.2034174002,  3.3582338303,  3.1268169264,\\  &3.1433784957,  3.2405135080, 3.1008649385,  3.2758823795,  3.1663607416,\\ &3.1524414836,  2.9931205862,  3.1318277759,  3.0787100614,  3.0011544184,\\ &2.9331935381,  3.0155597121,  3.0218345718,  2.9837460312,  2.9705769473,\\ &3.0755616247,  3.0868778008,  3.0677822108,  3.2140962589,  3.1436841099\\
    )
    \end{align*}

\end{document}